%% file: aied2019.tex
\documentclass[runningheads]{llncs}

\usepackage{amssymb}
\usepackage{graphicx}
\usepackage{url}
\usepackage{color}
\usepackage{amsfonts}
\usepackage{blindtext}
\usepackage{mathrsfs}
\usepackage{bbm}
\usepackage{url}
\usepackage{algorithm}
\usepackage{algorithmic}
\usepackage{paralist}
\usepackage[english]{babel}
\usepackage{booktabs}
\usepackage{multicol}
\usepackage{multirow}
\usepackage{subfig}

\begin{document}
\title{A Multimodal Alerting System for Online Class Quality Assurance}
%
%
\author{Jiahao Chen \and
Hang Li \and
Wenxin Wang \and
Wenbiao Ding \and
Gale Yan Huang \and
Zitao Liu\thanks{Corresponding Author: Zitao Liu}}

\authorrunning{J. Chen et al.}
%
\institute{TAL AI Lab, TAL Education Group, \\
DanLing SOHO, No.6 DanLing Street, Beijing, China \\
\email{\{chenjiahao, lihang4, wangwenxin2, dingwenbiao, galehuang, liuzitao\}@100tal.com}}
\maketitle              
\begin{abstract}
\input{abstract}
\end{abstract}

\section{Introduction}
\label{sec:intro}
\input{intro}

\section{The Multimodal Alerting System}
\label{sec:method}
\input{method}

\vspace{-0.3cm}
\section{Conclusion \& Future Work}
\vspace{-0.1cm}
\label{sec:conclusion}
\input{conclusion}

%
%
%
%
\bibliographystyle{splncs04.bst}
\bibliography{aied2019}
\end{document}

%% file: abstract.tex
Online 1 on 1 class is created for more personalized learning experience. It demands a large number of teaching resources, which are scarce in China. To alleviate this problem, we build a platform (marketplace), i.e., \emph{Dahai} to allow college students from top Chinese universities to register as part-time instructors for the online 1 on 1 classes. To warn the unqualified instructors and ensure the overall education quality, we build a monitoring and alerting system by utilizing multimodal information from the online environment. Our system mainly consists of two key components: banned word detector and class quality predictor. The system performance is demonstrated both offline and online. By conducting experimental evaluation of real-world online courses, we are able to achieve 74.3\% alerting accuracy in our production environment.
 

%% file: intro.tex
\begin{figure}%
\centering
\subfloat[Dahai online course scenario illustration. The virtual classroom consists of three panels: class materials, student and instructor.]{{\includegraphics[width=0.4\textwidth]{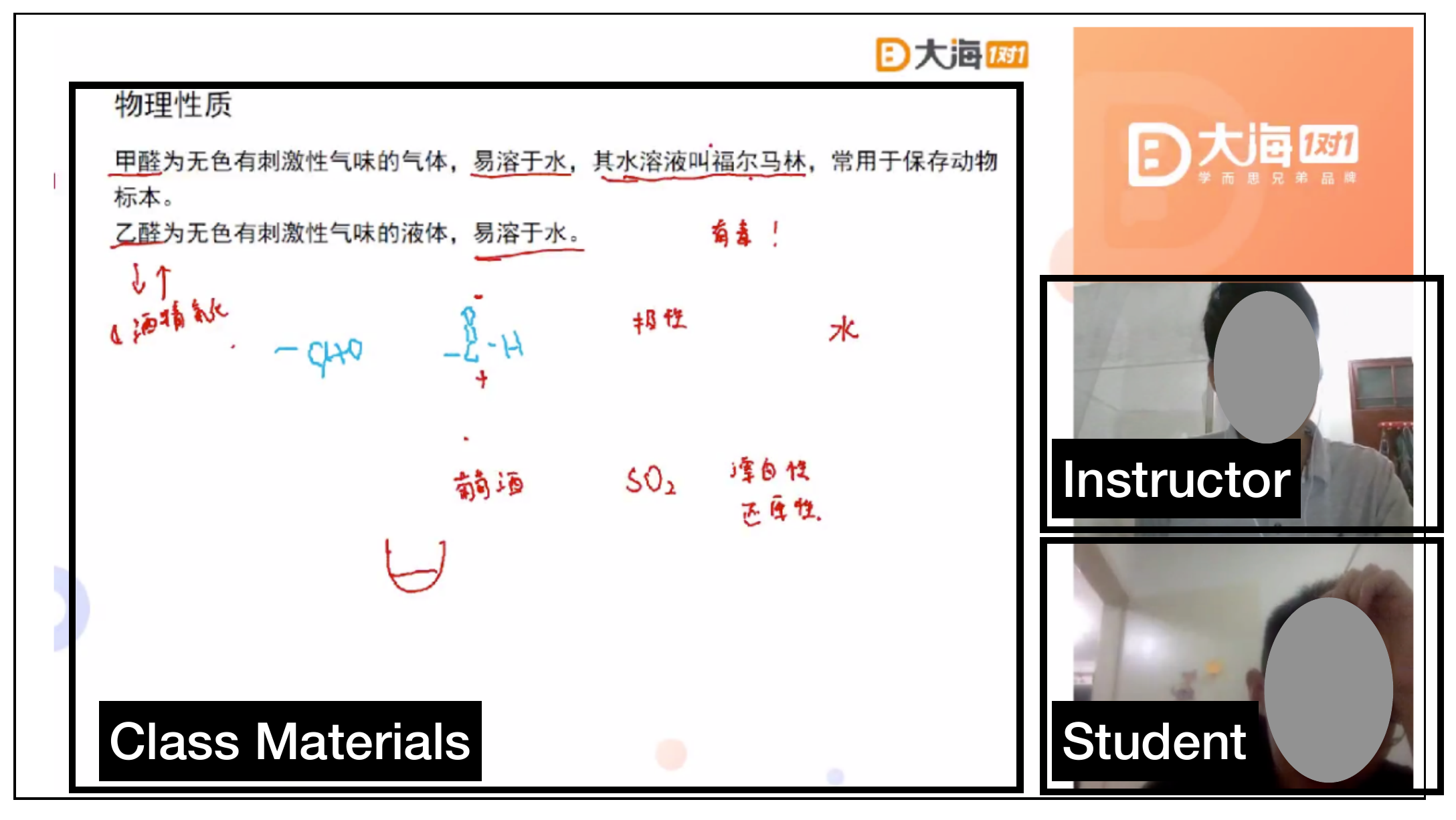} }}%
\qquad
\subfloat[Overview of our multimodal alerting system. The gray boxes indicate the key components in our system. ASR is short for automatic speech recognition.]{{\includegraphics[width=0.5\textwidth]{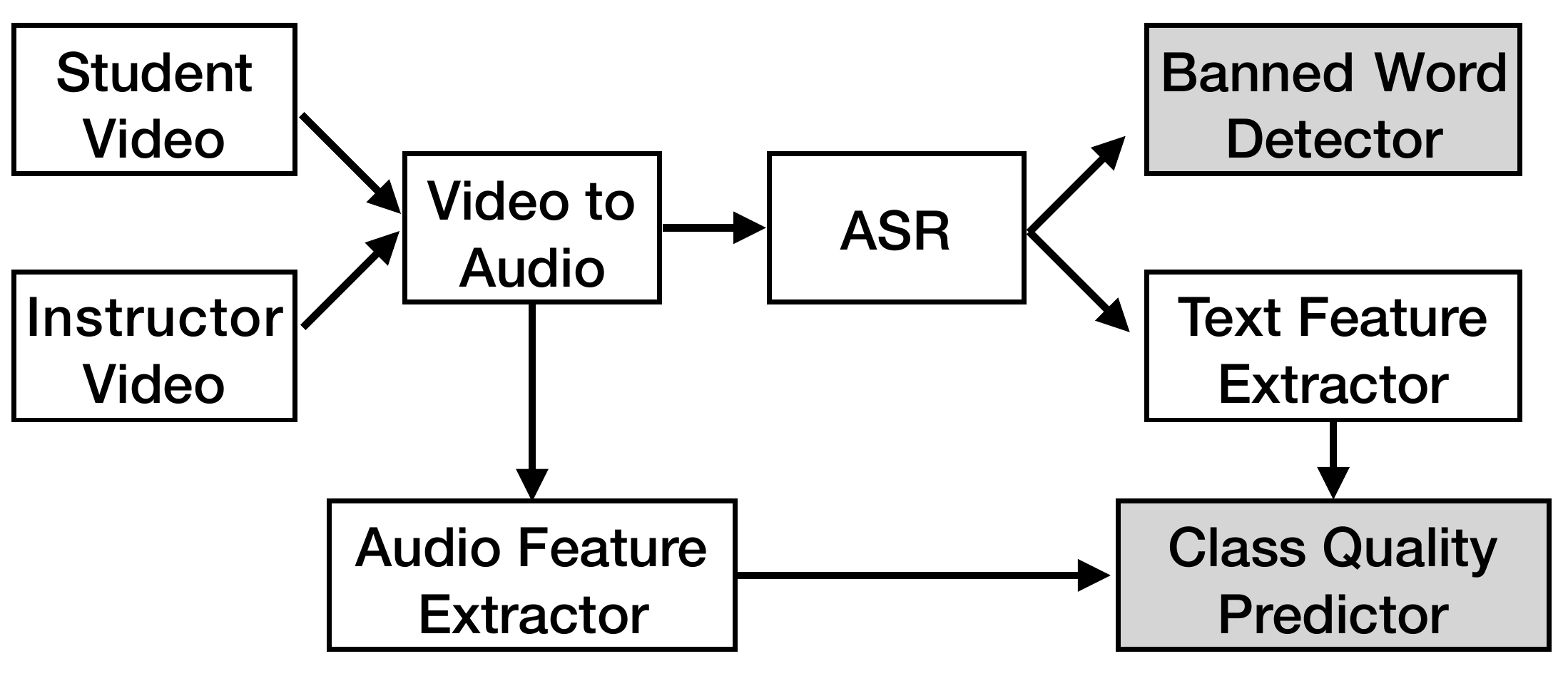} }}%
\caption{Dahai online course scenario illustration and an overview of our multimodal alerting system. Both student's and instructor's faces are hided by gray circles due to the privacy issue.}%
\label{fig:system}%
\end{figure}

With the recent development of technology such as digital video processing and live streaming, there has been a steady increase in the number of students enrolling online courses worldwide \cite{ChinaEducationResources2012}. Online 1 on 1 class is created to offer more personalized education experience. Both students and instructors are able to choose their out-class available time slots and have the class anywhere. To better allocate education resources in China, we create an online learning platform, i.e., \emph{Dahai} (\url{http://www.dahai.com}) with two distinct types of participants representing supply (instructors) and demand (students). On Dahai platform, instructors are senior college students from top Chinese universities and students come to Dahai for online tutoring. Once the study plan agreement is reached, the matched student and instructor start online courses in Dahai's virtual classroom via live streaming. Dahai  provides a wide range of online teaching tools to enable better teaching performance and interactions. Fig. \ref{fig:system}(a) shows the 1 on 1 learning environment provided by Dahai. Example industries include accommodation (Airbnb), ride sharing (Uber, Lyft, DiDi), online shops (Etsy, Taobao), etc. Without a doubt, quality assurance for these types of marketplaces need to satisfy both supply and demand sides of the ecosystem in order to grow and prosper \cite{abdallah2016fraud}. 

Allowing college students to be tutoring instructors\footnote{Tutoring instructors have to pass a series of interviews and training exercises before teaching the class.} is a double-edged sword. On one hand, it greatly alleviates the problem of imbalanced teaching resources in China. However, on the other, part-time instructors may not have enough teaching experience. Some unprofessional behaviors may lead to low class quality and inferior learning performance. Being an online education platform, Dahai is responsible for its class quality. The most common way to alleviate this problem is to allow students to give ratings for the online classes and detect low-quality classes by utilizing ratings. However, such approaches usually fail in online K-12 education since K-12 students rarely give responsible ratings. For example, a student may give a 5-star rating to an instructor teaching video games. Therefore, we build a multimodal alerting system to automatically monitor the quality of each class in Dahai. 

\vspace{-0.5cm}

\vspace{-1.2cm}

%% file: method.tex
\vspace{-0.3cm}

Multimodal information of the entire course is stored in the online learning environment. When a class is finished in Dahai, both the student and the instructor videos are passed to the backend for alerting and monitoring analysis. First, we extract audio tracks from videos. We transcribe the teaching conversations by using an automatic speech recognition (ASR) system. After that, we apply the banned word detector to scan all the contents to ensure there is no misbehavior happened. Second, we extract both the linguistic and prosodic features and build a logistic regression predictor to automatically evaluate the overall online 1 on 1 course quality. The entire workflow of our alerting system is illustrated in Fig. \ref{fig:system} (b).

\vspace{-0.3cm}
\subsection{Banned Word Detector}
\label{sec:keyword}

Instructors may thoughtlessly speak out swearing or insulting words or phrases. These words are referred to as \emph{banned words} and are definitely not allowed to appear in the class. However, banned words may happen in many scenarios. For example, instructors may lose their patience when students couldn't response after given many hints. Another example, instructors may speak their casual mantras during the class accidentally, which contains banned words. Besides, there are also cases that the teaching environment is very noisy and the banned words appear from the background. As an education platform, we must assure a cyberbully‐free and positive learning environment. Therefore, we develop the detector to take charge of banned word monitoring. We build the banned word detector by the following two steps:

\begin{itemize}
\item[Step 1.] We construct a banned word bank that covers all possible banned words and their variants. In Chinese, the smallest semantic unit is character instead of word. A word is made up of several characters. This leads to more linguistic variants. To tackle this problem, we first pre-define a seed set of banned words and then expand the seed set by finding the nearest neighbor words from the gigantic Chinese Internet corpus. The nearest neighbor search is conducted in the pre-trained Chinese word embedding space. The word embeddings are learned by directional skip-gram, which explicitly distinguishes left and right context in Chinese \cite{song2018directional}. After this expansion, we end up with a banned word bank with more than 3000 banned words. 
\item[Step 2.] We detect banned words by applying several fuzzy matching rules and heuristics. The matching procedure is challenging because of the recognition errors in the ASR transcriptions. To address this issue, we first write fuzzy regular expressions to retrieve banned word candidates. Our fuzzy regular expressions match not only the Chinese words but the romanization of the Chinese characters based on their pronunciation, i.e., Pinyin \cite{Wikipedia2019}. After that, we conduct Chinese word segmentation on the candidate words and their corresponding contexts. The segmentation process takes account into the semantic meaning of each candidate and eliminate false positive candidates. 
\end{itemize}

Here, we list a few classes caught by our detector in Table \ref{tab:example}. 

\vspace{-1cm}
\begin{table}
\centering
\caption{Examples of classes caught by the banned word detector.}
\label{tab:example}
\begin{tabular}{l|l}
\toprule
Examples & Instructor speech snippets with banned words \\
\midrule
Class\#1  & Read it again. \textbf{Fuck}, can't you remember these two sentence? \\
Class\#2  & \textbf{Damn it}. I knew it. You didn't read the paper. \\
Class\#3  & (Background noises) Come. Come. There is a group of \textbf{idiots}. \\
\bottomrule
\end{tabular}
\end{table}
\vspace{-1cm}

\subsection{Class Quality Predictor}
\label{sec:quality}

Besides catching the class with banned words, we are responsible for the overall quality of the class. The 1 on 1 online class is more like a black box that only happens between the instructor and the student. First, majority of parents have no time to watch their kids during the class, which makes no pressure from the demand side in this online marketplace. Second, students wouldn't tell the truth about the class quality. For example, we caught one class that the instructor spent the entire class talking about a mobile game, which makes the student highly satisfied. Third, one of the largest advantages of 1 on 1 class is that instructors are able to frequently interact with students. Students have many chances to ask questions and talk about their own thoughts. However, due to the lack of teaching experience, some instructors may still keep using the traditional offline teaching paradigm. There are barely any interactions and instructors talk for 60 minutes without stops.

Therefore, we build an automated quality predictor to monitor all the online courses on Dahai. We extract linguistic features from the ASR transcriptions and prosodic features from audio tracks. The linguistic features include the number of characters, words, and sentences, the number of class subject related words, etc. The prosodic features include signal energy, loudness, mel-frequency cepstral coefficients (MFCC) \cite{rabiner1975theory}, etc. We asked our teaching professionals to annotate 972 positive (good) courses and 219 negative (bad) courses. We use 80\% of them for training our logistic regression classifier and use the rest for testing purpose. We evaluate the effectiveness of linguistic and prosodic features respectively. We report accuracy, precision, recall and F1 score of the quality prediction performance in Table \ref{tab:prediction}. As we can see, both two types of features are very important to the quality prediction and the combination of both yields to the best results.

\vspace{-0.6cm}
\begin{table}
\centering
\caption{Offline experimental results of class quality prediction.}
\label{tab:prediction}
\begin{tabular}{l|c|c|c|c}
\toprule
Features &  Accuracy &  Precision  &  Recall  &  F1 score \\
\midrule
Linguistic Only & 0.897 & 0.899 & 0.986 & 0.940  \\
Prosodic Only & 0.949 & 0.944 & 0.997 & 0.970 \\
Linguistic + Prosodic  & 0.954 & 0.949 & 0.997 & 0.972  \\
\bottomrule
\end{tabular}
\end{table}
\vspace{-0.6cm}

\subsection{Online System Performance}

We deployed our monitoring and alerting system online. We set a few alerts based on the results of banned word detector and class quality predictor. Once the alarms are fired, we have operation staffs to watch the playback videos to conduct the final judgments. After comparing the staffs' ratings with our system's altering results, we achieve 74.3\% accuracy in system alerting.

%% file: conclusion.tex
In this paper, we presented our monitoring and alerting system for online 1 on 1 classes. By using the multimodal information, we are able to not only find misbehaviors in the online courses but measure the class quality. With the banned word detector and the class quality predictor, we are able to achieve 74.3\% accuracy in our online production system. In the future, we plan to explore information from the class materials panel and improve the alerting performance as well.